\documentstyle[11pt,epsfig]{article}
\begin{document}
\begin{titlepage}
\flushright{USM-TH-157}\\
\vskip 2cm
\begin{center}
{\Large \bf Effective action of dressed mean fields for  ${\cal N} =4$ 
super-Yang--Mills theory} \\
\vskip 1cm  
Gorazd Cveti\v{c}, Igor Kondrashuk, and Ivan Schmidt \\
\vskip 5mm  
{\it $(a)$ Departamento de F\'\i sica, 
Universidad T\'ecnica Federico Santa Mar\'\i a, \\ 
Avenida Espa\~{n}a 1680, Casilla 110-V, Valparaiso, Chile} \\
\end{center}
\vskip 20mm
\begin{abstract}
Based on general considerations such as $R$-operation and Slavnov-Taylor 
identity we show that the effective action, being 
understood as Legendre transform of 
the logarithm of the path integral, possesses particular structure
in  ${\cal N} =4$ supersymmetric 
Yang--Mills theory for kernels of 
the effective action expressed in terms of the dressed effective fields.
These dressed effective fields have been introduced in our previous papers 
as actual variables of the effective action. The concept  of dressed 
effective fields naturally appears in the framework of solution to 
Slavnov--Taylor identity. The particularity of the structure is 
the independence of these kernels on the ultraviolet regularization scale $\Lambda.$ 
These kernels are functions of mutual spacetime distances and of the gauge 
coupling. The fact that  $\beta$ function in this theory is zero 
is used significantly.
\vskip 1cm
\noindent Keywords: $R$-operation, gauge symmetry,  ${\cal N} =4$ supersymmetry,
Slavnov--Taylor identity. 
\end{abstract}
\end{titlepage}

\def\a{\alpha}
\def\b{\beta}
\def\db{{\dot{\beta}}}
\def\t{\theta}
\def\bt{{\bar{\theta}}}
\def\yb{{\bar{y}}}
\def\Db{{\bar{D}}}
\def\Tr{{\rm Tr}}
\def\dis{\displaystyle}
\def\le{\left(}
\def\ri{\right)}
\def\da{{\dot{\alpha}}}
\def\no{\nonumber}
\def\del{\delta}
\def\bW{\bar{W}}
\def\G{\Gamma}
\def\rar{\rightarrow}
\def\fra1g2{\frac{1}{g^2}}
\def\dg{{\dagger}}
\def\Vt{\tilde{V}}
\def\Kt{\tilde{K}}
\def\SQa{\sqrt{\tilde{\a}}}
\def\e{\epsilon}
\def\sm{{\sigma_m}}
\def\f12{\frac{1}{2}}
\def\F{\Phi}
\def\pd{\partial}
\def\ve{\varepsilon}
\def\F{\Phi}
\def\bF{\bar{\Phi}}

Slavnov--Taylor (ST) identity \cite{ST} is an important tool in quantum field theory.
It is a consequence of BRST symmetry \cite{BRST} of the tree level action 
of gauge theories, and it consists in an equation written for a functional 
that is called effective action \cite{SF}.
An approach to solving the ST identity in gauge theories has been proposed recently 
\cite{Cvetic:2002dx,Cvetic:2002in}. In this Letter 
we re-consider our analysis for a particular case of   ${\cal N} =4$ 
supersymmetric theory. Our analysis will be based on five theoretical tools:
$R$-operation \cite{BoSh}, gauge symmetry,  ${\cal N} =4$ supersymmetry,
the ST identity itself, and absorbing two point Green's functions into a
re-definition of the effective fields. Effective fields are variables 
of the effective action \cite{SF}.

${\cal N} =4$ super-Yang--Mills theory is widely considered  from the 
point of view AdS/CFT correspondence \cite{Maldacena:1997re}. 
Anomalous dimensions of gauge invariant 
operators are related to energies of string states \cite{Berenstein:2002jq}. 
In this Letter we consider 
${\cal N} =4$ super-Yang--Mills theory from a different point. We analyse 
one particle irreducible (proper) correlators of this theory, which 
are kernels of the effective action. For example, a kernel can be proper 
vertex of several gluons. We hope this 
analysis can have application to calculation of maximal helicity violating 
amplitudes of processes
with $n$ gluons \cite{Witten:2003nn,Bern:1994zx}.  
${\cal N} =4$ super-Yang--Mills theory is useful theoretical playground to 
understand better the problems that 
stand in QCD. This model has special particle contents. In addition to one gluon 
and four Majorana fermions it contains six scalar fields. All the 
particles are in the adjoint representation of $SU(N)$ gauge group.

It has been  known for a long time that for ${\cal N} =4$ Yang-Mills 
supersymmetric theory 
beta function of gauge coupling vanishes \cite{Ferrara:1974pu,Jones:1977zr,
Avdeev:1980bh,Sohnius:1981sn}. 
We extensively use 
this fact in our analysis. Basic notation of this Letter coincides with notation 
of Ref. \cite{Cvetic:2002dx}. The main ST identity is \cite{SF} 
\begin{eqnarray}
& \Tr\left[\dis{\int~d~x~\frac{\del \G}{\del A_m(x)}\frac{\del \G}{\del K_m(x)}
  + \int~d~x~\frac{\del \G}{\del c(x)}\frac{\del \G}{\del L(x)}
  - \int~d~x~\frac{\del \G}{\del b(x)}
   \le\frac{1}{\a}~\pd_m~A_m(x)\ri} \right] \no \\ 
& + \dis{~\int~d~x~\frac{\del \G}{\del \phi(x)}~\frac{\del \G}{\del k(x)} +
~\int~d~x~\frac{\del \G}{\del \bar{k}(x)}~\frac{\del \G}{\del\bar{\phi}(x)} = 0}.
\label{main} 
\end{eqnarray}
The effective action is a functional of all the effective fields and external sources
participating in this equation,  $\G \equiv 
\G[A_m,b,c,\phi,\bar{\phi},K_m,L,k,\bar{k}]$ \cite{Lee,SF}. The external sources 
$K_m,L,k,\bar{k}$ are coupled in the exponential of the path integral 
to the BRST transformations of fields from the measure of the path 
integral \cite{SF}, 
that is, to the BRST transformations of fields $A_m,c,\phi,\bar{\phi}$ 
of the tree level action, respectively.
The effective fields  $A_m,b,c,\phi,\bar{\phi}$ are traditionally named 
by the same letters as are  the fields  $A_m,b,c,\phi,\bar{\phi}$ which 
are variables of the path integral. The effective fields are defined as      
variational derivatives of the logarithm of the
path integral with respect to the corresponding external sources coupled to these
 variables
of integration in the path integral \cite{SF}.      
 The matter effective 
field $\phi$ stands for spinors as well as for scalars. We assume summations 
over all indices of the representation of matter fields. The
traditional Lorentz gauge fixing is taken and the corresponding Faddeev-Popov 
ghost action introduced 
according to the line of Ref. \cite{Avdeev:1980bh}.  These terms break supersymmetry 
of the tree level action. The $\b$ function is zero but anomalous dimensions 
of propagators 
are non-zero \cite{Avdeev:1980bh}.

Consider the vertex $Lcc$. Here we do not specify arguments 
of the effective fields. It is the only vertex which is invariant with 
respect to the ST identity at the classical level. At the quantum level 
it transforms to the form 
\begin{eqnarray}
\left<Lcc\right> \times \left<Lcc\right> + \left<LccA\right> 
\times \left<K_m\pd_mc\right> = 0.  \label{rough}
\end{eqnarray}
This is a direct consequence of the 
main ST identity (\ref{main}) and is a schematic form of the 
ST identity relating the $Lcc$ and $LccA$ field monomials. The precise form 
of this relation can be obtained by differentiating the identity (\ref{main}) 
with respect to $L$ and three times with respect to $c$ and then by  
setting all the variables of the effective action equal to zero. The brackets in   
(\ref{rough}) mean that we have taken functional derivatives with respect to 
fields in the corresponding brackets at different arguments and then have put all the 
effective fields equal to zero.

We know from the theory of $R$-operation \cite{BoSh} that in Yang-Mills theory 
the divergences  can 
be removed by re-defining the fields and the gauge coupling. Thus, there are four
renormalization constants that multiply  the ghost, gluon, spinor, and scalar 
fields \cite{SF}. The gauge coupling also must be renormalized but this is not 
the case in the 
theory under consideration.  In this paper we concentrate on two regularizations: 
regularization by higher derivatives described in Ref. \cite{SF}, $\Lambda$ is the 
regularization scale, and regularization by dimensional reduction.  
The regularization by higher derivatives  has been constructed 
for supersymmetric theories in Refs. \cite{Krivoshchekov:1978xg,West:1985jx}.  
Having used this regularization, new scheme has been proposed in 
Refs. \cite{Slavnov:2003cx,Slavnov:2002kg,Slavnov:2001pu}. We assume here 
that the component analog of that scheme can be constructed. 
The regularization by higher derivatives  provides strong suppression of ultraviolet 
divergences by introducing  additional terms with higher degrees of covariant 
derivatives acting on Yang--Mills tensor into the classical action, 
which are suppressed by appropriate degrees of the regularization scale 
$\Lambda.$  In addition to this, it is necessary to introduce a modification 
of the Pauli--Villars regularization to guarantee the convergence of the one-loop 
diagrams \cite{SF}. This scheme does not break gauge invariance beyond one loop 
level. Moreover, it has been suggested in Ref. \cite{SF} that
such a modification by Pauli--Villars terms to remove one-loop infinities 
is gauge invariant by construction. To regularize the fermion cycles, the usual 
Pauli--Villars regularization can be used. However, when applied to
explicit examples, this approach is known to yield incorrect results 
\cite{Martin:1994cg}. A number of proposals have been put forward to emend 
this problem \cite{Bakeyev:1996is}. 
However, all these proposals contain as intermediate steps some 
non-trivial extensions of the original setup. For instance they either require 
intermediate
dimensional regularization or include non-local terms in the action.
Because of these problems we analyse the theory in the regularization by the 
dimensional reduction in a parallel way.

At one-loop level 
the part associated with the divergence 
of $Lcc$ term must be invariant itself under the ST identity since the second term 
in identity (\ref{rough}) is finite in the limit of removing regularization 
$\Lambda \rar \infty$ \cite{Cvetic:2002dx}.  
According to Ref. \cite{Cvetic:2002dx}, this results in the following 
integral equation for the part of the correlator $Lcc$ corresponding  to the 
superficial divergence $\sim \ln\frac{p^2}{\Lambda^2}:$ 
\begin{eqnarray}
& \dis{\int~dx~\G_{\Lambda}(y',x,z')\G_{\Lambda} (x,y,z) = \int~dx~\G_{\Lambda} 
(y',y,x)\G_{\Lambda} (x,z,z')} \no\\
& = \dis{\int~dx~ \G_{\Lambda} (y',x,z)\G _{\Lambda} (x,z',y),}  \label{integral}
\end{eqnarray}
where $\G_{\Lambda} (x,y,z)$ is this scale($\Lambda-$) 
dependent part of the most general 
parametrization $\G (x,y,z)$ of the correlator $Lcc,$
\begin{eqnarray}
\dis{\G \sim \int~dx~dy~dz~\G (x,y,z)f^{abc}L^{a}(x)c^{b}(y)c^{c}(z).} 
\label{A1}
\end{eqnarray}
Here $f^{abc}$ is the group structure constant. The only solution to the integral 
equation (\ref{integral}) is \cite{Cvetic:2002dx}
\begin{eqnarray}
& \dis{\G_{\Lambda}(x,y,z) = \int~dx'~G_{c}(x'-x)~G^{-1}_{c}(x'-y)
     ~G^{-1}_{c}(x'-z).}        \label{result}
\end{eqnarray}
The subscript $\Lambda$ means scale-dependent part of the correlator.
As can be seen, all scale-dependence of this correlator is concentrated in 
the dressing function. The complete correlator $Lcc$ at one loop level can 
be then written as  
\begin{eqnarray*}
& \dis{\int~dx~dy~dz~\G (x,y,z)\frac{i}{2}f^{bca}
   L^{a}(x)c^{b}(y)c^{c}(z)} = \\
& = \dis{\int~dx'dy'dz'dxdydz~\tilde{\G} (x',y',z')G_{c}(x'-x)~G^{-1}_{c}(y'-y)}
\times \no\\
& \times~\dis{G^{-1}_{c}(z'-z)\frac{i}{2}f^{bca} L^{a}(x)c^{b}(y)c^{c}(z).} \no
\end{eqnarray*}
Here $\tilde{\G} (x',y',z')$ is scale-independent kernel of $Lcc$ correlator
\footnote{In Ref.\cite{Cvetic:2002dx}  we conjectured, based on Eqs.(\ref{integral}) 
and (\ref{A1}), that
the complete $\Gamma(x,y,z)$ has the structure (\ref{result}); 
this would correspond to 
$\tilde \Gamma(x',y',z') \propto \delta(x' - y') \delta(x' - z')$.}.

We absorb this dressing function $G_{c}$ 
into the corresponding re-definition of the fields $L$ and $c$, and then divide 
the ghost propagator in two parts one of which is related to the dressing 
function of the ghost field $G_c$ and another we call the dressing function of 
the gluon field $G_A$. The effective field $K$ and the antighost field $b$ 
get opposite re-definition 
by integrating with the dressing function $G^{-1}_{A}$ \cite{Cvetic:2002dx}. 
The important point here is 
covariance of the part of the ST identity without gauge fixing 
term with respect to such 
redefinitions \cite{Cvetic:2002dx}. In terms of the dressed effective 
fields we have a useful relation which is consequence of the main 
ST identity (\ref{main}) and can be obtained by differentiating  the main 
ST identity two times with respect to $\tilde{c}$ and one time with respect to 
$\tilde{b}.$ This resulting identity is 
\begin{eqnarray}
\left<\tilde{A}_m \tilde{b}\tilde{c}\right> \times \left<\tilde{K}_m 
\tilde{c}\right> + \left<\tilde{L}\tilde{c}\tilde{c}\right> 
\times \left<\tilde{b} \tilde{c}\right> = 0.  \label{rough2}
\end{eqnarray}
Again, this identity is written in a schematic way. However, a new important point 
appears here. Namely, since the two-point proper functions  in terms 
of dressed effective fields are trivial tree level two-point proper 
functions, the divergences of 
$\left<\tilde{A}_m \tilde{b}\tilde{c}\right>$ and 
$ \left<\tilde{L}\tilde{c}\tilde{c}\right>$ 
coincide. Since $ \left<\tilde{L}\tilde{c}\tilde{c}\right>$ is scale-(or $\Lambda-$)
independent, the $\left<\tilde{A}_m \tilde{b}\tilde{c}\right>$ is 
scale-independent also.  Concerning the gluon propagator,  one part of the divergence
is in the dressing function $G_A,$ and the rest of divergence would be  
absorbed in the re-definition of the gauge coupling constant. This last 
divergence is absent in ${\cal N} =4$ theories. All the other correlators 
are solved by the ST identity in terms of the dressed effective fields and 
their kernels are finite
(do not posses divergence in the limit of removing regularization) and 
scale-independent.

Infinite parts of the dressing functions will be one loop counterterms 
corresponding to 
the re-definition of the fields. Then we can repeat this procedure at two loop 
level and so on, up to any order in loop number.  Indeed, the re-definition 
by multiplication of the fields of the tree level action results in re-defining 
external legs of proper 
correlators in comparison 
with the unrenormalized theory. This property 
has been used by Bogoliubov and Shirkov \cite{BoSh} in the derivation 
of renormalization group equations. Re-defining 
the effective fields by dressing functions does not bring new aspects in 
this sense.
Indeed, re-defining variables of the path integral by dressing will result in 
the dressing of the external legs of the proper correlators.   
By proper correlators we mean kernels of $\G,$ that is, these kernels 
are one particle irreducible diagrams. 
We have already used 
this re-definition in the dressing functions in the $\tilde{L}\tilde{c}\tilde{c}$
correlator. Thus, new superficial 
divergences will have to satisfy the integral equation (\ref{integral}) at two 
loop order too in terms of effective fields  $\tilde{L}\tilde{c}\tilde{c}$ 
which are effective fields dressed by one loop dressing function $G_c.$ 
We then repeat this re-definition 
in the same procedure in each order of the perturbation theory as we did 
in the previous paragraph at one loop level.

Until now the pure gauge sector has been considered. Fermions are necessary 
for providing supersymmetry. Consider vertex $kc\phi$ at one loop level. 
The superficial divergence of this vertex is canceled by the divergence of 
the vertex $Lcc$. This means that in the part of the correlator  $kc\phi$ 
corresponding to superficial divergence  $\sim \ln\frac{p^2}{\Lambda^2}$ this 
divergence can be absorbed into the dressing 
functions in the following way:
\begin{eqnarray}
\int dxdx_1dy_1dy_2~G_{\phi}(x-x_1)~G^{-1}_{c}(x-y_1)
     ~G^{-1}_{\phi}(x-y_2)~k(x_1)~c(y_1)~\phi(y_2). 
   \label{ex4}
\end{eqnarray}
Note that the dressing functions for the fermions and scalars  $G^{-1}_{\phi}(x)$
are not fixed yet. We set them equal to halves of the two point matter 
functions. The rest of the vertices is restored in the unique way because 
the ST identity works.

This theory has intrinsic on-shell infrared divergences, like those canceled by  
brems\-stra\-hlung of soft gluons (such a cancellation happens on shell). 
To regularize these divergences we can introduce mass parameter $\mu$ 
\cite{Peskin}. Such a trick  breaks the ST identity by terms 
dependent on $\mu.$ At the end the dependence on $\mu$ will disappear
in physical matrix elements. We mean by the physical matrix element 
connected diagram on-shell contribution to amplitudes of particles. 
However, one can think that in the effective action (off shell) 
they could be  present since cancellation of the infrared divergences 
happens (on shell) between proper and one particle reducible graphs 
\cite{Peskin}. Below we will indicate that off shell these infrared divergences 
do not exist at all in the position space\footnote{We treat the theory 
in the position space. Infrared divergences are absent off shell also in 
the momentum space by the same reasons.}.

In principle, infrared divergences in the effective action represent an outstanding 
problem that is not treated in the present work in necessary details. We show in 
this paragraph that in the regularization by the dimensional reduction 
this problem does not appear. It is enough to show this for the $Lcc$ 
correlator, because other correlators can be expressed in terms of that one by ST 
identity if we work in terms of dressed effective fields. 
The one-loop contribution depicted in 
Fig. 1 (the only diagram that can be drawn) is apparently convergent 
in the Landau gauge. The point is that the derivatives 
can be integrated out of the graph due to the property of transversality of 
the gluon propagator in the Landau gauge what makes immediately this graph 
convergent in the ultraviolet region but is safe in the infrared. 
Note that in other gauges this correlator remains divergent in the ultraviolet, 
and its scale dependence 
is contained in the dressing function $G_c.$
\begin{figure}[ht]
\begin{center}
\epsfig{file=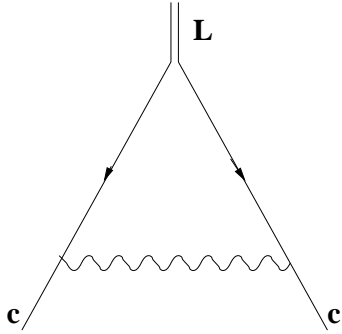, width=5.cm}
\end{center}
\vspace{0.0cm}
\caption{
\it One-loop contribution to the $Lcc$ vertex. The wavy lines
represent the gluons, the straight line are for the ghosts.}
\end{figure}
The two-loop diagrams (planar) are drawn below in Fig. 2. 
\begin{figure}[ht]
\begin{center}
\epsfig{file=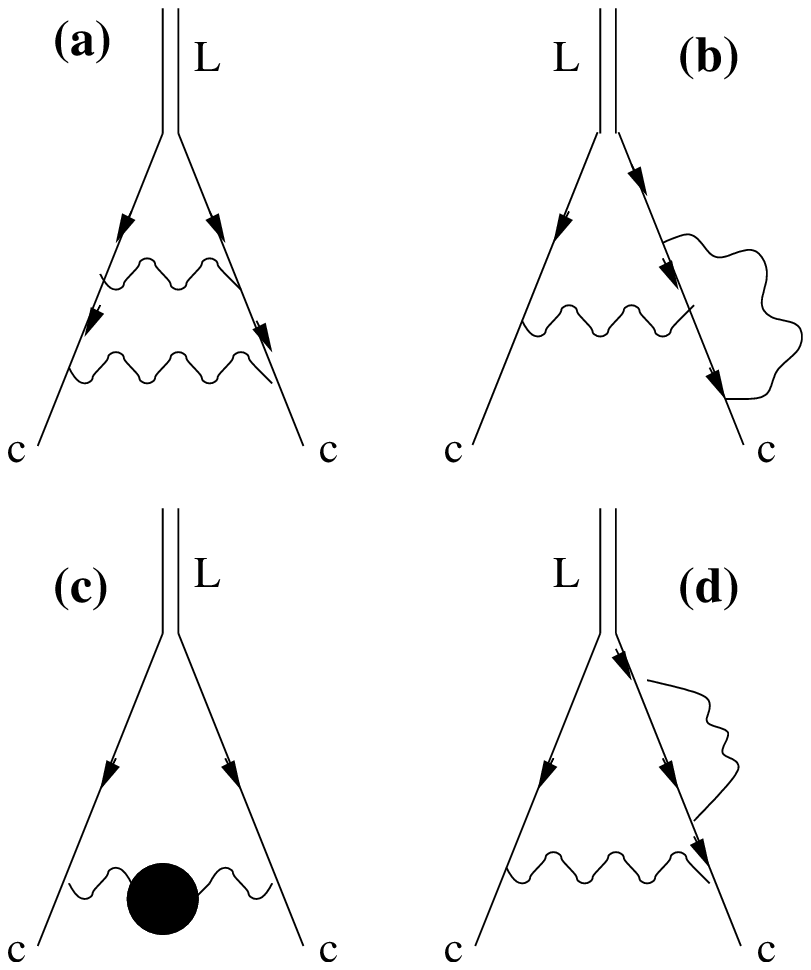, width=5.cm}
\end{center}
\vspace{0.0cm}
\caption{
\it  Two-loop diagrams for the $Lcc$ vertex. The wavy lines
represent the gluons, the straight lines the ghosts. The black disc 
in (c) is for one-loop contribution in the renormalization of the 
vector propagator from scalar, spinor and ghost fields}
\end{figure}
The first two diagrams are apparently convergent in ultraviolet in Landau 
gauge since all subgraphs are convergent. This is due to the property of 
transversality that allows to integrate out the derivatives again.
Infrared region is also not dangerous since even in Landau gauge the gluon 
propagator is safe in the infrared region.

The behavior of the theory in the IR region is not 
spoiled by the higher derivative regularization too. This is clear from the structure 
of the gluon propagator (in Landau gauge, for example) \cite{SF}:
\begin{eqnarray*}
\dis{D^{ab}_{\mu\nu} = \del^{ab}\left[-\le g_{\mu\nu} - \frac{k_\mu k_\nu}{k^2}
\ri\frac{1}{k^2 + k^6/\Lambda^4}\right].}
\end{eqnarray*} 
Sixth degree of momentum in the denominator improves 
significantly ultraviolet behavior but in the infrared it is negligible 
in comparison with 
second degree of momentum.  The infrared divergence appears  
when we en force put  the on-shell condition $p_i^2=0,$ where 
$p_i$ are external momenta. 
In general, infrared 
region is not dangerous off shell in component formulation in Wess-Zumino 
gauge when we regularize the theory by higher derivatives.

In such way we come to our main conclusion in this Letter. Namely, ${\cal N} =4$ 
supersymmetric 
theory has scale-independent effective action in terms of the dressed effective 
fields (once we disregard the possible dependence on the infrared 
regulator mass which is usually  taken to be infinitesimally small).  
All the dependence  on  the dimensionful parameter of ultraviolet 
regularization remains in the dressing 
functions only. This is in correspondence with direct calculation of anomalous 
dimensions and beta function in   ${\cal N} =4$ theory \cite{Avdeev:1980bh}. 
At one-loop level, kernels of this scale-independent theory are in general 
dilogarithms 
in momentum space. These dilogarithms are Fourier transforms of the kernels 
in position space as given below.  For example, the correlator of the 
dressed effective fields $\tilde{L},$
$\tilde{c},$ and $\tilde{c}$ at one loop level in any $SU(N)$ gauge theory 
has, among others, the following contribution:    
\begin{eqnarray}
\dis{\left<\tilde{L}^{a}(x)\tilde{c}^{b}(y)\tilde{c}^{c}(z)\right> \sim
g^2N \frac{1}{((z-y)^2)^2(x-y)^2(z-x)^2}f^{abc},} \label{structure}
\end{eqnarray}
where the dressed effective fields are made of
undressed effective fields convoluted to 
the dressing functions. The latter are unspecified but they are parts of the 
two point proper Green functions. The terms of the type (\ref{structure}) 
can be obtained, for example, by calculating the one-loop $Lcc$ Green function 
in the Landau gauge (where $G_c(x) = \del^{(4)}(x)$) and then using repeatedly 
the identity 
\begin{eqnarray*}
\frac{1}{(2\pi)^4}\int d^4k \frac{e^{-ikx}}{k^2 + i\e} = 
\lim_{\eta \rar +0} \frac{i}{4\pi^2} \frac{1}{\left[(|x^0| -i\eta)^2 - 
{\bf x}^2\right]}.
\end{eqnarray*}
By the ST identity the correlators 
(\ref{structure}) are related to the vertex $KAc$ which in its turn 
is related to the tree gluon vertex $AAA.$ The relation of these 
vertices is dictated by the ST identity and can be explicitly verified.      
Thus, the contribution similar to (\ref{structure}) can be found in the proper 
correlator of the three dressed gluons at one-loop level 
$<\tilde{A}_{\mu}^{a}(x)\tilde{A}_{\nu}^{b}(y)\tilde{A}_{\lambda}^{c}(z)>.$  
At the same time, in ${\cal N} =4$ supersymmetry we do not need to 
make additional renormalization in two point gluon function to 
absorb the rest of infinities from it into renormalization of the 
gauge coupling, since the $\b$ function is zero.

We found that the dressed effective fields are the actual variables of the 
effective action. The effective action is to be written in terms of these 
dressed effective fields. In general, in  non-supersymmetric gauge theory like QCD 
the dependence on UV regularization scale will be present inside the 
correlators of the dressed effective fields because it is necessary 
to remove the dependence on this scale by renormalization of the gauge 
coupling constant. In ${\cal N} =4$ supersymmetric theory such a renormalization 
does not take place. Thus, the kernels for the dressed fields do not 
depend on  scale.  This might make possible an analysis of 
these kernels by tools of conformal field theory in all orders of 
perturbation theory.

We have shown in this Letter that such a scale independent  structure of 
correlators is a direct consequence of the Slavnov--Taylor identity 
and it is encoded in the  $Lcc$ correlator of the dressed effective fields.
In general, by solving step-by-step ST identity it is possible to reproduce 
structure of all $n$-gluonic proper correlators in terms of dressed 
effective fields. Here the question how to define the concept of scattering 
can arise. The knowledge of correlators is not enough to define a scattering matrix. 
Indeed, it follows by the above construction that these correlators of dressed 
mean fields in ${\cal N} =4$ supersymmetry do not have dependence on any mass 
parameter, or, stated otherwise, the theory in terms of dressed mean fields 
is conformal invariant. It is known that $S$-matrix in conformal field theory 
cannot be constructed, since we do 
not have any dimensional parameter like mass or scale to define scattering concepts 
like typical scattering length, or size of a meson and so on. 
The only observables in this theory are correlators of the gauge invariant 
operators and their anomalous dimensions. However, on shell when we go to 
the amplitudes the scale appears due to infrared on shell divergences, 
so that scattering concepts can be introduced in a traditional way 
\cite{Bern:2005iz}.

\vskip 6mm
\noindent {\large{\bf{Acknowledgments}}}
\vskip 3mm

The work of I.K. was supported by Ministry of Education (Chile) under grant 
Mecesup FSM9901 and by DGIP UTFSM, and by  
Fondecyt (Chile) grant \#1040368. The work of G.C. and I.S. was supported by 
Fondecyt (Chile) grants \#1010094 and \#1030355, respectively.

\end{document}